# ASCA PV observations of the Seyfert 1 galaxy MCG−6−30−15 : rapid variability of the warm absorber.

C.S. Reynolds[1], A.C. Fabian[1], K. Nandra[1], H. Inoue[2], H. Kunieda[3] and K. Iwasawa[1,3]

[1] *Institute of Astronomy, Madingley Road, Cambridge CB3 0HA*
[2] *Institute of Space and Astronautical Science, Yoshinodai, Sagamihara, Kanagawa 229, Japan*
[3] *Department of Astrophysics, Nagoya University, Furo-cho, Chikusa, Nagoya 464, Japan*

14 June 1995

**ABSTRACT**
We present a detailed re-analysis of the two *ASCA* Performance Verification observations of the nearby Seyfert 1 galaxy MCG−6−30−15. Confirming the results of Fabian et al. (1994), we find definite evidence for the O VII and O VIII K-shell absorption edges of the warm absorber and a doubling of the warm absorber column density within the 3 weeks separating the two observations. No intra-day *flux-correlated* variability of the warm absorber is found. However, we report the discovery of an 'event' in which the warm absorber parameters temporarily change for $\sim 10\,000$ s before returning to their original values. Possible interpretations are discussed but a contradiction remains: the constancy of the ionization state of the warm absorber argues that it lies at large distances from the central source whereas the short term change in column density argues for small distances. Fluorescent iron emission is examined. As found by Fabian et al. (1994), the iron line is broad and strong (equivalent width $\sim 300$ eV). The line profile is also suggestive of it being skewed. Such a line would be expected from a relativistic accretion disk.

We also find very rapid primary X-ray variability. Assuming relativistic beaming to be unimportant, the derived efficiency is comparable to the maximum obtainable from accretion onto a Schwarzschild black hole. Correlated variability outside of the energy range of *ASCA* might exceed this maximum, thus requiring efficient accretion onto a Kerr hole.

## 1 INTRODUCTION

The nearby bright Seyfert 1 galaxy MCG−6−30−15 has been extensively studied in X-rays. A short *EXOSAT* observation revealed X-ray variability on timescales of 1 500 s (Pounds, Turner & Warwick 1986), making it one of the first active galactic nuclei (AGN) in which such rapid variability was known. Further *EXOSAT* observations showed evidence for the fluorescent K-shell emission and absorption of cold iron (Nandra et al. 1989) expected from the illumination of dense, cold material located near the central engine by the primary X-rays (Guilbert & Rees 1988; Lightman & White 1988). Confirmation of these cold iron features by *Ginga* as well as the discovery of the associated reflected continuum supported this picture (Nandra, Pounds & Stewart 1990; Matsuoka et al. 1990). Nandra, Pounds & Stewart (1990) also found evidence for partially-ionized, optically-thin gas along the line of sight from the iron K-edge. This material has become known as the warm absorber (Halpern 1984). Supporting evidence for the warm absorber model was provided by *ROSAT* Position Sensitive Proportional Counter (PSPC) observations which suggested the presence of absorption edges due to O VII and O VIII at 0.74 keV and 0.87 keV respectively (Nandra & Pounds 1992).

*ASCA* (Tanaka et al. 1994) observed MCG−6−30−15 twice during the Performance Verification (PV) phase of the mission. These observations are reported by Fabian et al. (1994), hereafter F94. They clearly resolved O VII and O VIII K-shell edges for the first time. In addition, they found that the iron K-line at 6.4 keV was broad. This paper presents a detailed analysis of the *ASCA* PV data for MCG−6−30−15. Soft ($< 2$ keV) spectral features and variability are considered within the framework of warm absorber models. Our examination of the hard spectral features concentrates on the fluorescent cold iron K-line.

2  *C. S. Reynolds et al.*

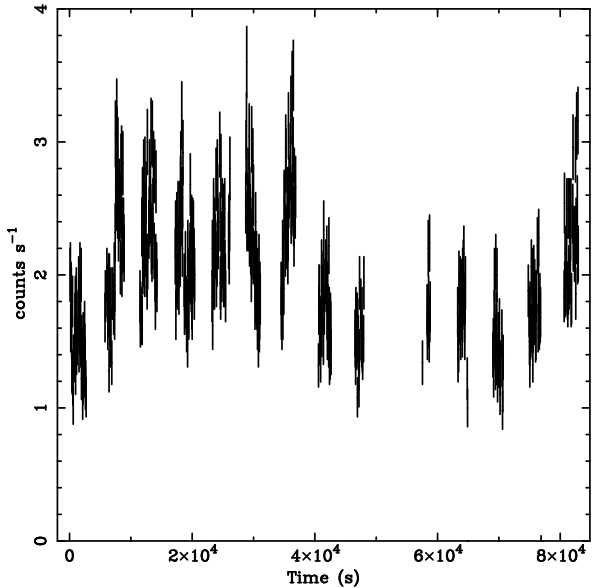

**Figure 1.** SIS0 light curve (0.5–10 keV) for the July dataset with 50 s bins. This light curve displays dramatic variations which are also seen in the other detectors.

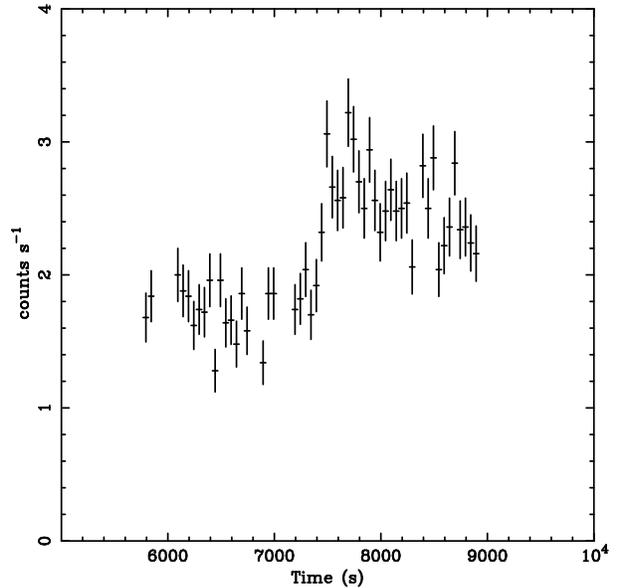

**Figure 2.** SIS0 light curve (0.5–10 keV) of the period between 5 000 s and 10 000 s after the beginning of the observation. This shows one of the most dramatic changes in luminosity in which the SIS0 count rate increases by a factor of 1.5 in ∼100 s. Similar behaviour is seen in the other detectors. As in Fig. 1, 50 s bins are used.

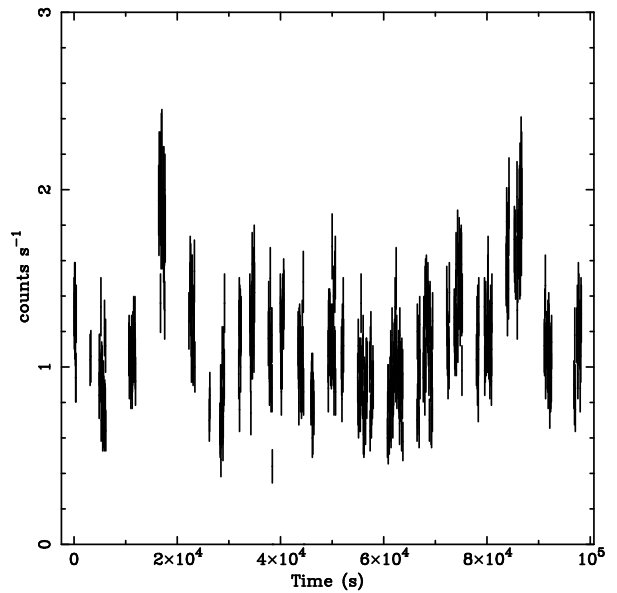

**Figure 3.** SIS0 light curve (0.5–10 keV) for the August dataset. 50 s bins are used. Several strong flares are seen with the luminosity doubling and then falling back to its original value within ∼10 000 s.

Section 2 briefly describes the observations and demonstrates the rapid variability exhibited by this object. Section 3 examines the soft spectral features and variability within the framework of warm absorber models. Implications of the observations on the location and origin of the material are discussed. This is of wide importance: the study of these ionized absorbers is the first case in which a varying photoionizing continuum and the responding absorption edges are simultaneously and directly observed. Section 4 investigates the iron line feature, including its variability. Section 5 provides a summary.

## 2 OBSERVATIONS

MCG−6−30−15 was observed by *ASCA* on 1993 July 8–9 and 1993 July 31–August 1 (hereafter referred to as the July and August datasets). Each of these two observations extend over one day and have ∼30 000 s of good data per detector (once contamination from the Earth and the South Atlantic Anomaly has been removed). Unless otherwise stated, Solid-state Imaging Spectrometer (SIS) data taken in both Faint and Bright mode have been combined.

The SIS0 0.5–10 keV light curve for the July dataset is shown in Fig. 1. The average 2–10 keV flux for this dataset was $4.6 \times 10^{-11}$ erg cm$^{-2}$ s$^{-1}$. The corresponding luminosity is $1.2 \times 10^{43}$ erg s$^{-1}$ (assuming $H_0 = 50$ km s$^{-1}$ Mpc$^{-1}$). Several rapid changes in luminosity can be seen. In the most dramatic event (Fig. 2), the 0.5–10 keV luminosity increases by 50 per cent (from ∼ $1.8 \times 10^{43}$ erg s$^{-1}$ to ∼ $2.8 \times 10^{43}$ erg s$^{-1}$) in ∼100 s. The change in luminosity over the time taken for the change is

$$\frac{\Delta L}{\Delta t} \approx 10^{41} \text{ erg s}^{-2}$$

Assuming this is a true representation of the intrinsic variability (i.e. there are no effects from relativistic beaming), the efficiency of conversion of accreted mass into radiative energy (Fabian 1979) is at least 5 per cent. This is comparable to the maximum efficiency obtainable from accretion onto a Schwarzschild black hole (Shapiro & Teukolsky 1983). This behaviour is also seen in the other SIS (SIS1) and the two Gas Imaging Spectrometers (GIS2 & GIS3). Correlated flux variations at energies outside of the *ASCA* band will serve to increase this lower limit to the efficiency. Thus, the efficiency may well exceed the maximum obtainable from a Schwarzschild black hole indicating the presence of highly efficient accretion onto a Kerr hole.

Fig. 3 shows the SIS0 light curve for the August dataset. The average 2–10 keV flux for this dataset is $3.9 \times 10^{-11}\,\mathrm{erg\,cm^{-2}\,s^{-1}}$. The corresponding 2–10 keV luminosity is $1.1 \times 10^{43}\,\mathrm{erg\,s^{-1}}$. This lacks the dramatic short term variation of the July dataset but does demonstrate strong 'flares' in which the luminosity doubles and then returns to the quiescent value during a period of $\sim 10\,000\,\mathrm{s}$.

## 3 THE WARM ABSORBER

F94 confirmed the presence of partially-ionized optically-thin material in the line of sight to the primary X-ray source. This has a large effect on the soft (0.5–2 keV) spectrum of MCG–6–30–15, with absorption edges of O VII and O VIII being the dominant features (see Fig. 1 of F94). They also found evidence for a doubling of the column density of warm material between the two PV observations. Here we perform a more detailed time resolved study of the warm absorber. We examine variability within each dataset as well as changes between the datasets.

### 3.1 Modelling the warm absorber

Following F94, the photoionization code CLOUDY (Ferland 1991) was used to examine the passage of the primary X-rays through an optically-thin, geometrically-thin shell of material. CLOUDY solves the equations of thermal and ionization equilibrium self-consistently with the radiation field. The incident spectrum is assumed to be a power law extending between 13.6 eV and 40 keV with photon index $\Gamma$. We also constrain the density to be fixed at $n = 10^{9.5}\,\mathrm{cm^{-3}}$, although the results are very insensitive to the actual density chosen (this is true for densities between $n \sim 10\,\mathrm{cm^{-3}}$ and $n \sim 10^{11}\,\mathrm{cm^{-3}}$). Such models can be parameterized by the column density, $N_\mathrm{W}$, and ionization parameter[*], $\xi$, of the material. These models are one-zone models in the sense that the density and ionization parameter of the absorbing gas is uniform.

A three-parameter grid of such models was computed in which $N_\mathrm{W}$, $\xi$ and $\Gamma$, are free parameters. These models can be compared with data using the XSPEC ATABLE utility. Such a procedure gives the best fit values for $N_\mathrm{W}$, $\xi$ and $\Gamma$.

---

[*] The ionization parameter, $\xi$, is defined by
$$\xi = \frac{L}{nR^2}$$
where $L$ is the incident ionizing luminosity, $n$ is the electron number density and $R$ is the distance of the gas from the source of ionizing luminosity. This form for the ionization parameter is chosen due to the ease with which it can be related to observation.

| model parameter | July dataset | August dataset |
|---|---|---|
| $\Gamma$ | $1.98 \pm 0.01$ | $1.87 \pm 0.01$ |
| $\tau_{\mathrm{OVII}}$ | $0.53 \pm 0.04$ | $0.63 \pm 0.05$ |
| $\tau_{\mathrm{OVIII}}$ | $0.19 \pm 0.03$ | $0.44 \pm 0.04$ |
| $\chi^2/\mathrm{dof}$ | 679/470 | 758/470 |

**Table 1.** Results of power-law and O VII/O VIII edge fit to the July and August datasets. Data from SIS0 and SIS1 were fitted simultaneously (with different instrumental normalisations) to obtain these results. $\Gamma$ is the photon index of the primary power-law. $\tau_{\mathrm{OVII}}$ and $\tau_{\mathrm{OVIII}}$ are the optical depths at threshold of the O VII and O VIII K-shell absorption edges respectively (with the edge energies held fixed at the physical values). The errors quoted are at the 90 per cent confidence level, $\Delta \chi^2 = 2.7$.

A major limitation of these models is the restriction to conditions of thermal and ionization equilibrium. Indeed, the data presented here lead us to suggest that both thermal and ionization equilibria may not be satisfied. Krolik & Kriss (1995) have recently examined models of X-ray heated winds with the assumption of thermal equilibrium relaxed. They show that the relaxation of this assumption has an important influence on the resulting emission, reflection and absorption. However, the current equilibrium models do provide a useful parameterisation of the data, provided they are interpreted carefully.

### 3.2 Comparison of the July and August datasets

In order to study spectral changes between the July and August datasets, a single integrated spectrum was extracted from each dataset for each of the four instruments on *ASCA*. The spectra were background subtracted using source free regions of the same observation. Initially, the data were fitted with a simple model consisting of a power law primary spectrum (photon index $\Gamma$), cold absorption (fixed at the Galactic value of $N_\mathrm{H} \approx 4.1 \times 10^{20}\,\mathrm{cm^{-2}}$; Elvis, Wilkes & Lockman 1989) and the K-shell absorption edges of O VII and O VIII (with the edge energies held fixed at the physical values of 0.74 keV and 0.87 keV respectively). [Throughout this work, energies of spectral features (e.g. absorption edges and emission lines) are quoted in the source rest-frame (with $z = 0.008$).] The free-parameters in such a fit are $\Gamma$, the optical depths of the edges and the absolute normalisation. Data from the two SIS detectors were fitted simultaneously. The results from these fits are presented in Table 1. It can be seen that both edges undergo an increase in optical depth, although the change is more dramatic for the O VIII edge. $\Gamma$ also decreases by $\sim 0.1$.

Self-consistent warm absorber models (as described in Section 3.1) were fitted simultaneously to all four instruments (leaving the normalisation of each instrument free to allow for the slight discrepancy between the four instruments). Fig. 4 shows the July dataset fitted with such a model. When performing such fits, cold absorption is included with a column density fixed at the Galactic value. Table 2 gives the results of such fits to the two datasets. As can be seen from Fig. 4, the major discrepancies between these warm absorber models and the data occur in the range 5–7 keV and may be interpreted in terms of K-shell emission of iron. Details of this feature are to be investigated in Section 4. It is relevant to the warm absorber analysis only in so far as it could bias the fit towards lower values of $\Gamma$. To eliminate any such possible biasing, we performed a second fit including only these data below 4 keV (also shown in Table 2). Excluding data above 4 keV leads to a steepening of the inferred photon index and a slight decrease in $\xi$ for the warm absorbing material.

| model parameter | July (full spectrum) | August (full spectrum) | July (< 4 keV) | August (< 4 keV) |
|---|---|---|---|---|
| $L_{\rm abs}$(0.5–2 keV) | 0.8 | 0.5 | 0.8 | 0.5 |
| $L_{\rm abs}$(2–10 keV) | 1.2 | 1.1 | 1.1 | 0.9 |
| $L_{\rm unabs}$(0.5–2 keV) | 1.1 | 0.8 | 1.2 | 0.9 |
| $L_{\rm unabs}$(2–10 keV) | 1.2 | 1.1 | 1.1 | 1.0 |
| $\log N_{\rm W}$ | 21.78±0.03 | 22.07±0.03 | $21.79^{+0.02}_{-0.03}$ | $22.03^{+0.02}_{-0.03}$ |
| $\xi$ | 32±2 | $43^{+3}_{-2}$ | 26±2 | 34±3 |
| $\Gamma$ | 1.97±0.02 | 1.87±0.02 | 2.06±0.02 | 2.00±0.02 |
| $\chi^2$/dof | 1732/1390 | 1637/1330 | 926/825 | 822/766 |

**Table 2.** Results of three-parameter warm absorber fits to integrated July and August datasets. $L_{\rm abs}$ is the observed luminosity (after reprocessing by the warm absorber) whereas $L_{\rm unabs}$ is the inferred luminosity prior to warm absorption. All luminosities are given in units of $10^{43}\,{\rm erg\,s^{-1}}$. $N_{\rm W}$ is given in units of $\rm cm^{-2}$ and $\xi$ is in units of $\rm erg\,cm\,s^{-1}$. The errors are quoted at the 90 per cent confidence level, $\Delta\chi^2 = 2.7$. Fits which ignore data above 4 keV are free of any possible biasing introduced by the iron line.

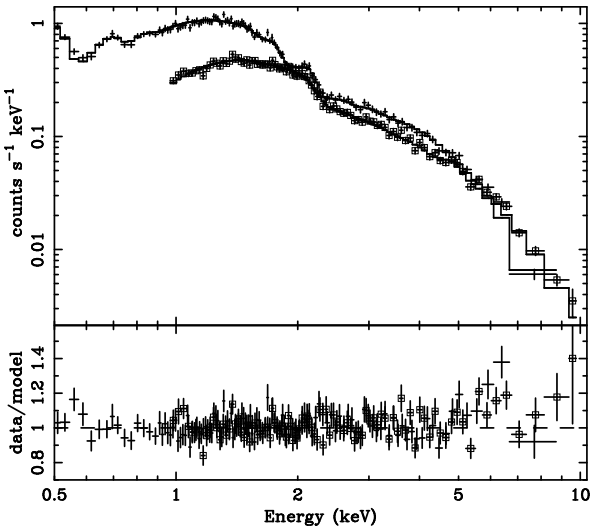

**Figure 4.** Integrated (background subtracted) spectrum for the July dataset. All four instruments are used in the spectral fitting but, for clarity of presentation, only SIS0 (plain crosses) and GIS2 (crosses with squares) are shown in the figure. Also shown is the best fit warm absorber model.

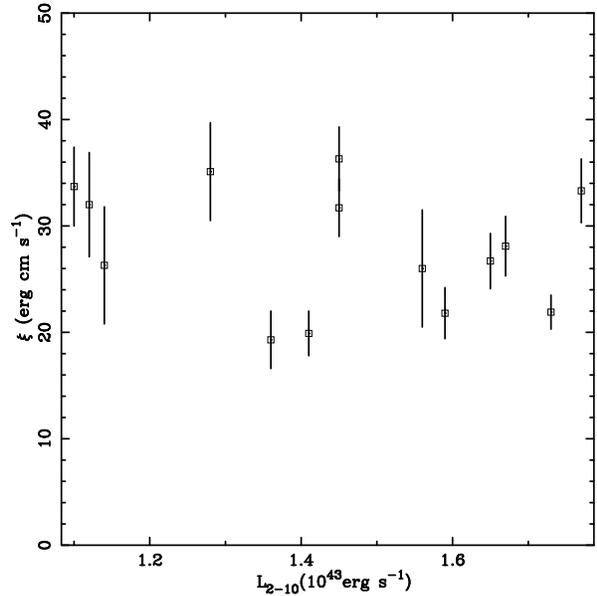

**Figure 5.** Inferred $\xi$ plotted against the 2–10 keV luminosity for the July dataset. $N_{\rm W}$ and $\Gamma$ are held constant at $6\times10^{21}\,{\rm cm^{-2}}$ and 2, respectively. Errors are given at the 90 per cent confidence level for one interesting parameter ($\Delta\chi^2 = 2.7$).

Comparing either the two full spectra or the two truncated spectra leads to the conclusion that the warm column density has increased from $(6.2\pm0.4)\times10^{21}\,{\rm cm^{-2}}$ to $(1.1\pm0.1)\times10^{22}\,{\rm cm^{-2}}$, i.e. the column density has doubled in three weeks. There is also evidence for an increase in $\xi$ between the July and August datasets despite a significant decrease in the ionizing luminosity. These results confirm the findings of F94. Comparison of the full spectra shows a decrease of $\Gamma$ by 0.1. However, this effect is much less significant in the truncated spectra and thus could be due to the behaviour of the hard spectrum. Some implications of these results are discussed presently.

### 3.3 Short term variability

The reaction of the warm absorbing material to the rapid fluctuations in the primary ionizing luminosity is an important diagnostic of the physical conditions within this gas.

| model parameter | July (first half) | July (second half) | August (first half) | August (second half) |
|---|---|---|---|---|
| $L_{\rm abs}(0.5\text{--}2\,{\rm keV})$ | 0.9 | 0.7 | 0.5 | 0.6 |
| $L_{\rm abs}(2\text{--}10\,{\rm keV})$ | 1.3 | 1.0 | 0.9 | 1.0 |
| $L_{\rm unabs}(0.5\text{--}2\,{\rm keV})$ | 1.3 | 1.0 | 0.9 | 1.0 |
| $L_{\rm unabs}(2\text{--}10\,{\rm keV})$ | 1.4 | 1.0 | 1.0 | 1.2 |
| $\log N_{\rm W}$ | $21.75\pm0.03$ | $21.89\pm0.06$ | $22.01\pm0.03$ | $22.04^{+0.03}_{-0.04}$ |
| $\xi$ | $23\pm2$ | $30^{+3}_{-5}$ | $33\pm2$ | $35\pm2$ |
| $\Gamma$ | $2.09^{+0.04}_{-0.03}$ | $2.06\pm0.04$ | $2.02\pm0.03$ | $2.00\pm0.03$ |
| $\chi^2/{\rm dof}$ | 735/765 | 583/570 | 674/675 | 681/680 |

**Table 3.** Results of three-parameter warm absorber fits to each half of the July and August datasets. $L_{\rm abs}$ is the observed luminosity (after reprocessing by the warm absorber) whereas $L_{\rm unabs}$ is the inferred luminosity prior to warm absorption. All luminosities are given in units of $10^{43}\,{\rm erg\,s^{-1}}$. $N_{\rm H}$ is given in units of ${\rm cm^{-2}}$ and $\xi$ is given in units of ${\rm erg\,cm\,s^{-1}}$. Spectra for all four instruments were fitted simultaneously. Data above 4 keV were ignored to prevent contamination by iron emission. The errors are quoted at the 90 per cent confidence level, $\Delta\chi^2 = 2.7$.

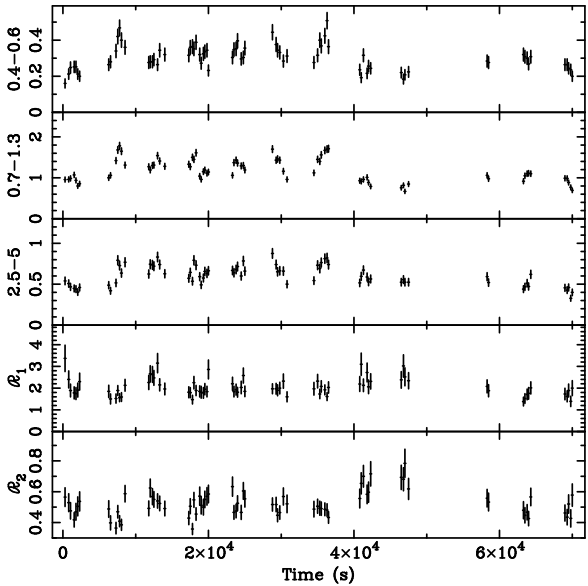

**Figure 6.** Summed SIS0 and SIS1 light curve in various bands for the July dataset. The panels show (from top to bottom) a) the 0.4–0.6 keV count rate, b) the 0.7–1.3 keV count rate, c) 2.5–5 keV count rate, d) ratio of 2.5–5 keV data to 0.4–0.6 keV data (hardness ratio $\mathcal{R}_1$) and e) ratio of 2.5–5 keV data to 0.7–1.3 keV data (hardness ratio $\mathcal{R}_2$). All count rates are in counts s$^{-1}$. 250 s bins were used. Note that during the period between 40 000 s and 50 000 s $\mathcal{R}_2$ suffers a significant rise with no associated strong change in $\mathcal{R}_1$. This is evidence for a rapid change in the warm absorber.

Limits can be set on the photoionization and recombination timescales of the material thereby allowing the density to be constrained. Constraints on the density together with measured values for $\xi$ and $N_{\rm W}$ give information on the distance of the material from the central ionizing source (Reynolds & Fabian 1994). In addition, inhomogeneities can be studied.

Here we examine each dataset for short term variation of the warm absorber. Two techniques were used to analyse the short term variability of the warm absorber; variation of colour ratios and direct fitting of warm absorber models to spectra taken from various times within the dataset.

### 3.3.1 The July dataset

The July dataset was divided into two equal halves. Spectra were extracted for each half and warm absorber models fitted to the data below 4 keV. Table 3 shows the results of such fits. There is evidence for an increase in both the column density and ionization parameter of the warm material between the two halves of the dataset, despite a significant decrease in ionizing flux. The power-law index of the primary spectrum, $\Gamma$, does not change significantly.

The dataset was divided further. Spectra were extracted for each spacecraft orbit (giving 14 spectra for each instrument). Warm absorber models were then fitted to each of these spectra. Unfortunately, there were insufficient counts to usefully constrain three-parameter warm absorber models. The major statistical degeneracy is a trade-off between $N_{\rm W}$ and $\xi$. To counter this degeneracy, we repeated the warm absorber fits holding $N_{\rm W}$ and $\Gamma$ fixed at $6\times 10^{21}\,{\rm cm^{-2}}$ and 2, respectively. The instrument normalisations were also tied together in an attempt to reduce statistical degeneracies. Thus, $\xi$ and the overall model normalisation were the only free parameters in such fits. Fig. 5 shows the results of such fits. Here, we plot the inferred $\xi$ as a function of the 2–10 keV luminosity (taken to be indicative of the total ionizing luminosity). Within this parameterisation, significant changes of $\xi$ are seen. However, they are uncorrelated with ionizing luminosity and may be unrelated to the warm absorber. We can rule out the hypothesis of constant $N_{\rm W}$, constant $\Gamma$ and $\xi$ proportional to the ionizing flux.

Changing colour ratios are another indicator of varying warm absorbers (Netzer, Turner & George 1994). Warm absorption predominately affects the flux in the 0.7–1.3 keV range. On the other hand, 2.5–5 keV flux and 0.4–0.6 keV flux should be largely unaffected by warm absorption or iron emission. Thus, we define two hardness ratios

$$\mathcal{R}_1 = \frac{2.5 - 5\,{\rm keV\ count\ rate}}{0.4 - 0.6\,{\rm keV\ count\ rate}} \quad (1)$$

$$\mathcal{R}_2 = \frac{2.5 - 5\,{\rm keV\ count\ rate}}{0.7 - 1.3\,{\rm keV\ count\ rate}} \quad (2)$$

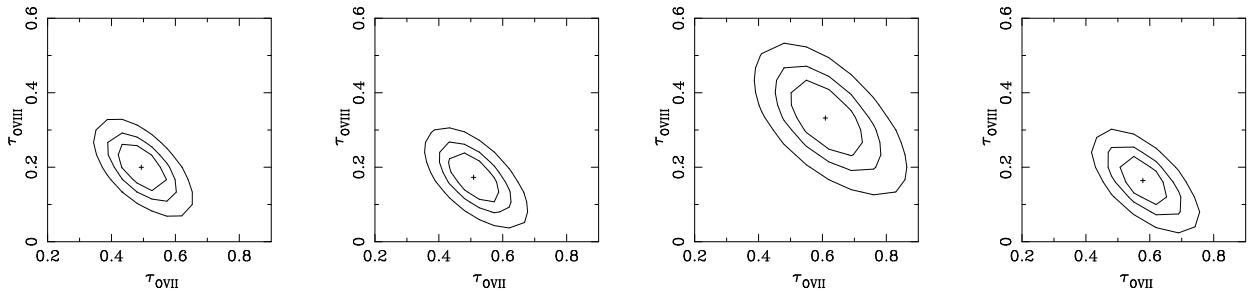

**Figure 7.** Confidence contours for the optical depths at threshold of the O VII and O VIII edges ($\tau_{\rm OVII}$ and $\tau_{\rm OVIII}$ respectively) during spectral fitting with the power-law and 2 edge model. (a), (b), (c) and (d) show orbits 4/5, 6/7, 8/9 and 10+ respectively. The contours show the 68 per cent, 90 per cent and 99 per cent confidence regions for two interesting parameters.

For a fixed primary spectral shape, changes in the parameters of the warm absorber will lead to a change in $\mathcal{R}_2$ whilst leaving $\mathcal{R}_1$ unchanged. Fig. 6 shows the 0.4–0.6 keV count rate, the 0.7–13 keV count rate, the 2.5–5 keV count rate and these two hardness ratios.

Fig 6 shows that for a period between 40 000 s and 50 000 s after the beginning of the observation (i.e. spacecraft orbits 8 & 9 of the observation), $\mathcal{R}_2$ undergoes a significant increase with little corresponding change in $\mathcal{R}_1$. This 'hardening event' was investigated further. SIS spectra were accumulated for the data taken during this period. Comparison spectra were accumulated for two periods before this event (orbits 4/5 and 6/7) and one period after the event (orbits 10+). Table 4 shows the results of various spectral fits to these spectra. Fig. 7 shows the confidence contours for the optical depths at threshold of the O VII and O VIII edges ($\tau_{\rm OVII}$ and $\tau_{\rm OVIII}$ respectively) in the power-law and 2 edge fit. It is clear that the warm absorption undergoes a change (leading predominantly to an increase in $\tau_{\rm OVIII}$) and then returns to its original state.

#### 3.3.2 *The August dataset*

As with the July dataset, the August dataset was divided into two equal halves, spectra were extracted for each half and warm absorber models were fitted to each half of the dataset. Table 3 reports the results of these fits. There was no significant change in any of the warm absorber parameters between the two halves of this dataset (although the average luminosity does not change dramatically between these two halves of the dataset). As with the July dataset, spectra for individual spacecraft orbits do not contain enough counts to constrain three-parameter warm absorber models. Fixing $N_{\rm W}$ and $\Gamma$ at $1.2 \times 10^{22}$ cm$^{-2}$ and 1.9 leads to no significant variation in $\xi$ between the fits.

Fig. 8 shows the 0.4–0.6 keV count rate, the 0.7–1.3 keV count rate, the 2.5–5 keV count rate and the hardness ratios $\mathcal{R}_1$ and $\mathcal{R}_2$. Both hardness ratios are consistent with being constant over the period of the August observation (constant model gives $\chi^2$/dof=75/74 and 81/74 for $\mathcal{R}_1$ and $\mathcal{R}_2$ respectively). Thus the data are consistent with the hypothesis that the warm absorber remains unchanged over the duration of the August dataset.

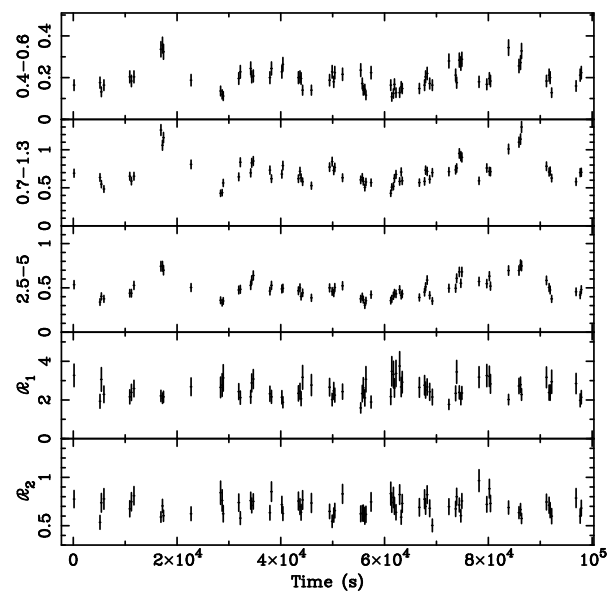

**Figure 8.** Summed SIS0 and SIS1 light curve in various bands for the August dataset. The panels show (from top to bottom) a) the 0.4–0.6 keV count rate, b) the 0.7–1.3 keV count rate, c) 2.5–5 keV count rate, d) ratio of 2.5–5 keV data to 0.4–0.6 keV data (hardness ratio $\mathcal{R}_1$) and e) ratio of 2.5–5 keV data to 0.7–1.3 keV data (hardness ratio $\mathcal{R}_2$). All count rates are in counts s$^{-1}$. 250 s bins were used. No significant changes in hardness ratios are seen.

### 3.4 Discussion

#### 3.4.1 *Summary of the observational results*

As reported by F94, there is a clear change in the characteristics of the warm absorption between the July and August datasets. The dominant change is an increase in $\tau_{\rm OVIII}$, although $\tau_{\rm OVII}$ may also have increased. This accompanies a significant decrease in ionizing luminosity. One-zone warm absorber models interpret this as a doubling of $N_{\rm W}$ and a small increase in $\xi$.

The July dataset shows evidence for short term warm absorber variability. In an event lasting $\sim 10\,000$ s, a slight decrease in ionizing luminosity is accompanied by a significant increase in $\tau_{\rm OVIII}$. Again, one-zone warm absorber models interpret this as a doubling of $N_{\rm W}$. The warm absorber parameters then return to their pre-event values. This change is very similar to the change seen between the July and August datasets suggesting a possible bi-modal behaviour. We note that there are similar decreases in luminosity which are *not* accompanied by a change in the warm absorber. No significant change in the warm absorption is seen in the August dataset.

| model parameter | Orbits 4 & 5 | Orbits 6 & 7 | Orbits 8 & 9 | Orbits 10+ |
|---|---|---|---|---|
| $L_{\rm abs}$(0.5–2 keV) | 0.9 | 1.0 | 0.6 | 0.7 |
| $L_{\rm abs}$(2–10 keV) | 1.4 | 1.5 | 1.3 | 1.0 |
| $L_{\rm unabs}$(0.5–2 keV) | 1.1 | 1.4 | 1.0 | 0.9 |
| $L_{\rm unabs}$(2–10 keV) | 1.4 | 1.5 | 1.4 | 1.1 |
| $\Gamma$ | $2.02 \pm 0.02$ | $2.03 \pm 0.02$ | $1.94 \pm 0.04$ | $2.03 \pm 0.03$ |
| $\tau_{\rm OVII}$ | $0.49 \pm 0.07$ | $0.51 \pm 0.09$ | $0.60 \pm 0.14$ | $0.58 \pm 0.09$ |
| $\tau_{\rm OVIII}$ | $0.20 \pm 0.07$ | $0.17 \pm 0.08$ | $0.32 \pm 0.11$ | $0.17 \pm 0.07$ |
| $\chi^2$/dof | 232/244 | 282/232 | 204/201 | 270/220 |
| $\log N_{\rm W}$ | $21.76 \pm 0.08$ | $21.64^{+0.07}_{-0.08}$ | $21.86^{+0.08}_{-0.10}$ | $21.71^{+0.05}_{-0.07}$ |
| $\xi$ | $30 \pm 7$ | $19^{+5}_{-7}$ | $28^{+5}_{-8}$ | $21 \pm 4$ |
| $\Gamma$ | $2.02 \pm 0.03$ | $2.08^{+0.08}_{-0.05}$ | $1.98^{+0.08}_{-0.04}$ | $2.08 \pm 0.04$ |
| $\chi^2$/dof | 255/248 | 282/232 | 200/201 | 278/220 |

**Table 4.** Results of spectral fitting to data taken from orbits 4/5, 6/7, 8/9 and 10+ of the July dataset. Two separate models were fitted to each period of data; a simple power-law plus O VII/O VIII edge model (as in Table 1) and three-parameter warm absorber models (as described in Section 3.1). Only SIS data below 4 keV was used.

### 3.4.2 Important timescales

Initially, we make two assumptions; that the warm absorber is one-zone (in the sense of Section 3.1) and that the ionization state of the material is dominated by photoionization by the primary radiation field. Relaxation of these assumptions will be discussed below. Given these assumptions, there are at least three characteristic timescales relevant to a discussion of warm absorber variability; the variability timescale of the primary ionizing flux ($t_{\rm var}$), the recombination timescale of the dominant ions within the warm material ($t_{\rm rec}$) and the photoionization timescale of the dominant ions within the material ($t_{\rm ph}$). The recombination timescale (to all atomic levels) for O VIII is given approximately by

$$t_{\rm rec} \approx 200 n_9^{-1} T_5^{0.5} \, {\rm s}$$

where $n = 10^9 n_9 \, {\rm cm}^{-3}$ is the number density of the warm material and $T = 10^5 T_5$ K is the temperature of the warm material (Allen 1973; Shull & van Steenberg 1982). Using the definition of $\xi$, this can be expressed as

$$t_{\rm rec} \approx 200 \xi_2 R_{16}^2 L_{43}^{-1} T_5^{0.5} \, {\rm s}$$

where $\xi = 10^2 \xi_2 \, {\rm erg \, cm \, s}^{-1}$, $R = 10^{16} R_{16}$ cm is the distance of the material from an source with ionizing luminosity $L = 10^{43} L_{43} \, {\rm erg \, s}^{-1}$.

We can also provide a simple estimate of $t_{\rm ph}$. If we assume the ionizing continuum has a power-law form with photon index $\Gamma = 2$ extending between $E_{\rm min}$ and $E_{\rm max}$, and that the K-shell photoionization cross-section is proportional to $\nu^{-3}$ (above the threshold energy), the K-shell photoionization timescale is given by

$$t_{\rm ph} \approx \frac{16\pi\Lambda R^2 E_{\rm th}}{\sigma_{\rm th} L}$$

where $\sigma_{\rm th}$ is the photoionization cross-section at the threshold energy $E_{\rm th}$ and

$$\Lambda = \ln\left(\frac{E_{\rm max}}{E_{\rm min}}\right)$$

For O VII, this evaluates to

$$t_{\rm ph} \approx 20 \, R_{16}^2 L_{43}^{-1} \, {\rm s}$$

Thus, $t_{\rm ph} < t_{\rm rec}$ provided $\xi_2 T_5^{0.5} > 0.1$.

If $t_{\rm var}$ is longer than the other relevant timescales, the material would achieve photoionization equilibrium. The observationally inferred $\xi$ would be proportion to the ionizing flux. The present data rule out this scenario. We note that in at least one *Ginga* observation the inferred ionization parameter *does* vary in proportional to the flux at 1 keV (see Fig. 7 of Nandra, Pounds & Stewart 1990). However, these *Ginga* fits may be sensitive to other forms of spectral variability such as changes in $\Gamma$ or changes in the amount of reflection.

More generally, the ionization parameter $\xi$ would be expected to respond to increases in the ionizing flux on the photoionization timescale, whereas $\xi$ would respond to decreases in the ionizing flux on the recombination timescale. The observed ionization state would then depend on a weighted average of the past flux history.

### 3.4.3 Implications for warm absorber: beyond the one-zone model

The present data show no apparent *flux-correlated* changes in $\xi$ over a period of a day but an increase in $\xi$ over the three week period. [Note that the apparent discrepancy between the change in $\xi$ and the change in the ionizing flux (i.e. $\xi$ shows an increase whereas the ionizing flux decreases) can be reconciled if the ionizing flux flared up during the unobserved period.] Thus, we can conclude that both $t_{\rm rec}$ and $t_{\rm ph}$ are longer than a day, but that $t_{\rm ph}$ is shorter than (or of the order of) 3 weeks. These constraints on the timescales can be converted into constraints on the distance of the material from the ionizing source. Thus, on the basis of the above one-zone models, we would place the material between $\sim 6 \times 10^{17} L_{43}$ cm and $\sim 3 \times 10^{18} L_{43}$ cm. This is thought to be somewhat outside the broad line region (BLR). However, we *do* observe an event in which the warm absorber parameters undergo a change in 10 000 s or less. Within the one-zone models, this would argue that $t_{\rm ph} < 10\,000\,s$ and thus that the material lies within $\sim 2 \times 10^{17} L_{43}$ cm. Therefore, the one-zone models produce a clear contradiction.

Realistic models for this material may require the consideration of inhomogeneities or stratification within the material, the possibility that the ionization state is not dominated by photoionization and the possibility that the material is far from equilibrium. We now qualitatively discuss these some of these complications.

Inhomogeneities in the warm material would have an important effect on variability. Such inhomogeneities could arise from multiphase behaviour of the material leading to the formation of warm clouds. Multiphase behaviour would be particularly relevant for models in which the warm material was considered as a component of the emission line regions (see the discussion in Reynolds and Fabian 1994). Suppose the 10 000 s 'hardening event' seen in the July dataset corresponds to the passage of a warm cloud in front of the primary X-ray source. Light travel arguments together with the observed 100 s primary variability of this object imply a size of $\sim 10^{13}$ cm for the X-ray emission region. To traverse this region in 10 000 s, the warm cloud must have a velocity greater than $\sim 10^9\,{\rm cm\,s^{-1}}$. This places the warm material coincident with, or within, the BLR (i.e. $\sim 10^{15}$ cm) provided gravitational forces dominate.

Until now, we have been treating the primary X-ray source as a varying point source. Realistically, the X-ray source will have a finite extension. For example, primary X-rays may originate from the inner accretion disk (and hence have an extent of $10^{13}$–$10^{14}$ cm) with isolated 'flares' producing the variability on timescales down to $\sim 100$ s. If the warm absorber is inhomogeneous on length scales of $10^{13}$–$10^{14}$ cm, the properties of the absorbing gas along the line of sight to different flaring region could be different. This might lead to complicated and chaotic changes in the fitted warm absorber parameters. If the X-ray emission were dominated by only two flaring regions, the bi-modality of the warm absorber seen in the present *ASCA* datasets could have a natural explanation since two lines of sight are involved. Within this scenario, recombination and photoionization timescales would no longer dominate a discussion of inferred warm absorber variability.

Alternatively, stratification could be important. As an illustration of a simple stratified model, consider the situation in which the warm absorber consisted of two spatially distinct regions; a hot inner region responsible for some portion of the O VIII absorption edge and a cooler outer region responsible for the O VII absorption edge and some part of the O VIII absorption edge. Suppose the outer region has long photoionization and recombination timescales (longer than 3 weeks) whereas the inner region has a recombination timescale shorter than three weeks. The small drop in flux between the two PV observations could cause the hot, inner region to partially recombine, therefore causing the optical depth of the O VIII edge to increase without an associated change in the O VII edge (which is compatible with the observed changes noted in Section 3.2). When fit with a one-zone warm absorber model, we would infer an increase in $\xi$ and $N_{\rm W}$ (since the O VIII/O VII ratio would have increased and the total absorption would have increased). However, this model has difficulty in explaining the short term variability of the warm absorber unless we invoke an inhomogeneous hot region. This example shows the one-zone warm absorber parameterisation could lead to misleading results.

Finally, we discuss the possibility that photoionization does not dominate the ionization state of the material. The presence of an additional heat source for the warm material would lead to an equilibrium state lying to the left of the photoionization equilibrium curve on Fig. 3 of Reynolds & Fabian (1995). Mechanical heating driven by the primary radiation pressure is one such possible heat source. In this case, the ionization state of the material would be determined mainly by collisional effects. The response of the ionization state and column density of the warm material to variability in the primary flux would depend on the response of the mechanical heating to the flux variability and *not* directly on the primary ionizing flux. Optical depth changes in the absorber might lead to feedback processes operating on the mechanical heating. Thus the variability of the state of the absorber would be complicated, including the possibility of hystersis effects and instabilities in the absorbing material. We note that the opposite case, in which heating exceeds radiative cooling, is examined by Krolik & Kriss (1995).

## 4  THE IRON LINE

MCG$-$6$-$30$-$15 shows clear evidence for K-shell emission from iron, as discussed by F94. The reflection of X-rays from optically-thick, cold material (e.g. the accretion disk or the molecular torus) would result in the 6.4 keV fluorescent line of iron. In principle, the profile of this line is an important diagnostic of the conditions within the central engine. If this line originates from reflection by an accretion disk around a supermassive black hole, it is expected to have a characteristic asymmetric double peak structure resulting from the effects of Doppler shifting, relativistic beaming and gravitational reddening. Alternatively, if the line originates from reflection by the molecular torus that is proposed to exist in the standard model of Seyfert galaxies, we would expect a narrow line due to the small velocity dispersion of the torus material. The variability of this emission in response to primary flux variations also gives important information on the location of the line emitting material. The line flux variations would expected to follow the primary X-ray flux variations but would be lagged and smeared by the characteristic light travel time of the line emitting region.

Here we examine the profile and variability of this emission feature. All spectral fitting in this section includes background subtraction (with background taken from source free regions within the same observation).

### 4.1  Spectral Analysis

Spectral fits to the iron emission feature were found to be extremely sensitive to the parameterisation of the underlying

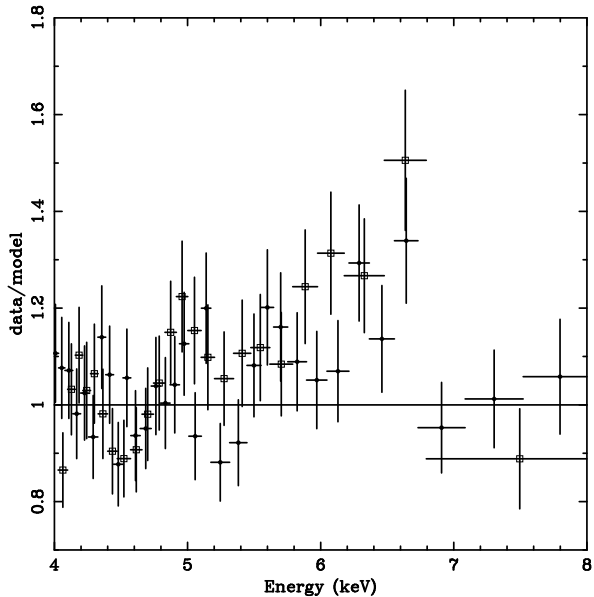

**Figure 9.** Ratio of the July data to a model consisting of a power-law (including the effects of reflection) and two absorption edges (to model the warm absorber). The open squares are SIS0 data (Faint mode data only) and the filled dots are GIS2 data (note that the spectral fitting was performed using all four instruments simultaneously but we only show two detectors for clarity of presentation). The iron emission feature can be clearly seen to be broad.

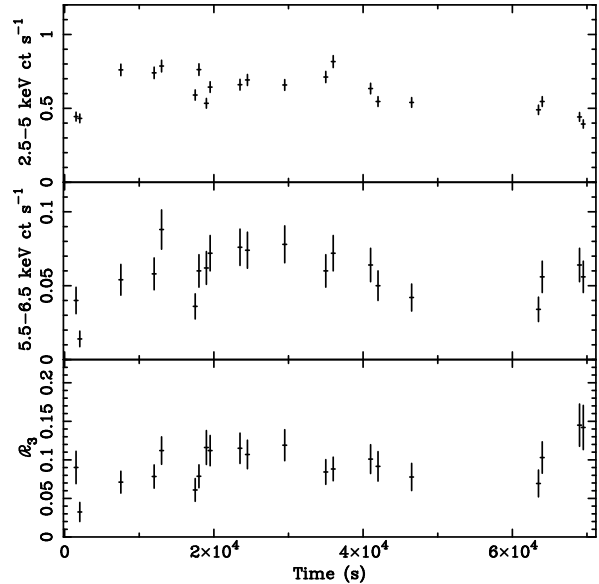

**Figure 10.** 2.5–5 keV count rate, 5.5–6.5 keV count rate and the hard to soft ratio of these two count rates ($\mathcal{R}_3$) for the July dataset. SIS0 and SIS1 data have been summed to produce these light curves. 500 s bins are used.

continuum. To eliminate as many systematic errors as possible, Bright mode SIS data was discarded for the purposes of detailed spectral fitting (i.e. only Faint mode SIS data and GIS data was used). This reduces the amount of good SIS data to ∼20 000 s for the July dataset and ∼10 000 s for the August dataset (per SIS detector). Hence the July dataset will provide better statistical constraints than the August dataset [Note that F94 use only Bright mode data and thus obtain better constraints from the August dataset]

Initially, each dataset was fitted with a power law and two absorption edges at low energies (to model the warm absorber). The resultant fits give $\chi^2$/dof of 1704/1658 and 1386/1423 for the July and August datasets respectively. As noted in Section 3, the dominant residuals in such a fit are between 5 keV and 7 keV and are attributed to iron emission. Fig. 9 shows the ratio of the July data to this model. This suggests the presence of a broad skew emission line. The addition of a Gaussian emission line proves to be highly significant (with $\Delta\chi^2$ of 73 and 59 for the July and August datasets respectively.) For the July dataset, the line energy is $E = 6.2 \pm 0.2$ keV, the line width is $\sigma = 0.7 \pm 0.2$ keV and the equivalent width is EW=400 ± 100 eV. The August dataset gives $E = 5.9 \pm 0.3$ keV, $\sigma = 0.8^{+0.4}_{-0.2}$ keV and EW=$500^{+200}_{-120}$ eV.

*Ginga* spectra of MCG−6−30−15 show spectral flattening at high energies, interpreted as the presence of a reflected continuum superposed on the primary power-law. Incorporating such a reflected continuum significantly affects the results. The datasets have insufficient counts to constrain the parameters of the reflection (such as solid angle subtended by the reflecting material at the X-ray source or inclination of the reflecting material). Thus, we fixed the reflection to be face-on with the reflecting material subtending $2\pi$ steradians at the primary X-ray source (i.e. as expected if the reflection were from an accretion disk illuminated by an isotropic X-ray source lying above the plane of the disk). Fitting a reflected power-law (and two low energy edges to model the warm absorber) to each dataset results in $\chi^2$/dof of 1673/1658 and 1358/1424 for the July and August datasets respectively. This improves the fit over the simple (unreflected) power-law by $\Delta\chi^2$ of 31 and 34 respectively. The addition of a Gaussian emission line again leads to a further significant improvement with $\Delta\chi^2$ of 36 and 28 for the two datasets respectively. The line parameters for the July data are $E = 6.2 \pm 0.3$ keV, $\sigma = 0.6^{+0.3}_{-0.2}$ keV and EW=260 ± 100 eV. The August dataset gives $E = 5.9^{+0.2}_{-0.3}$ keV, $\sigma = 0.7^{+0.3}_{-0.2}$ keV and EW=300 ± 100 eV.

### 4.2 Short Term Variability

Due to the small count rate within this emission feature, the only method to examine short term variability of the iron emission is via colour ratios. On the basis of ratio plots such as Fig. 9, we would expect somewhat more than 20 per cent of the 5.5–6.5 keV counts to originate from the iron emission. Thus we define the hardness ratio

$$\mathcal{R}_3 = \frac{5.5 - 6.5 \text{ keV count rate}}{2.5 - 5 \text{ keV count rate}} \qquad (3)$$

In particular, we seek to test the hypothesis that the line is constant during large changes in primary flux. Assuming the fluorescence is driven by the primary flux, this would give information on the size of the emitting region.

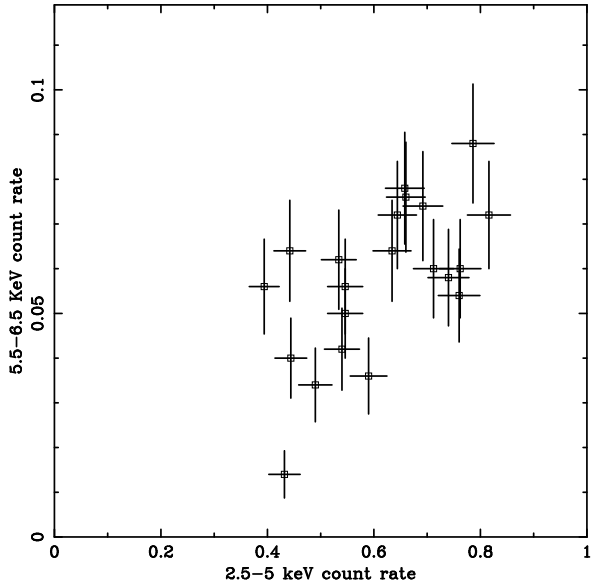

**Figure 11.** 5.5-6.5 keV count rate as a function of 2.5–5 keV count rate for the July dataset. SIS0 and SIS1 data have been summed. Each point represents 500 s of data.

*4.2.1  The July dataset*

Fig. 10 shows the 2.5–5 keV count rate, 5.5–6.5 keV count rate and $\mathcal{R}_3$, as a function of time for the July dataset. The hypothesis of a constant $\mathcal{R}_3$ is unacceptable with $\chi^2/\text{dof}=46/21$. Fig. 11 shows the same data with the 5.5–6.5 keV count rate shown as a function of the 2.5–5 count rate. We can constrain any constant component in the 5.5–6.5 keV count rate to be less than 0.001 counts s$^{-1}$ using a constant plus linear parameterisation of this relation. This sets an upper limit to the equivalent width of any constant iron emission to be $\sim 10$ eV (at the 90 per cent confidence level, $\Delta\chi^2 = 2.7$). However, this result weakens significantly if the point with the lowest 5.5–6.5 keV count rate is excluded. The upper limit on the constant component then becomes 0.04 counts s$^{-1}$ corresponding to an upper limit for the equivalent width of any constant iron emission to be $\sim 250$ eV.

*4.2.2  The August dataset*

Fig. 12 shows the 2.5–5 keV count rate, 5.5-6.5 keV count rate and $\mathcal{R}_3$ as a function of time for the August dataset. A striking feature of these light curves is the constancy of the colour ratio during the significant flare which occurs 20 000 s into the observation. The ratio is consistent (at the 90 per cent level) with a constant ($\chi^2/\text{dof}=14/19$; best fit ratio of $0.097\pm0.008$). Fig. 13 shows the 5.5-6.5 keV count rate as a function of the 2.5–5 keV count rate. We can constrain any constant component in the 5.5-6.5 keV count rate to be less than 0.02 counts s$^{-1}$, corresponding to an upper limit of $\sim 300$ eV to the equivalent width for any constant line (at the 90 per cent confidence level).

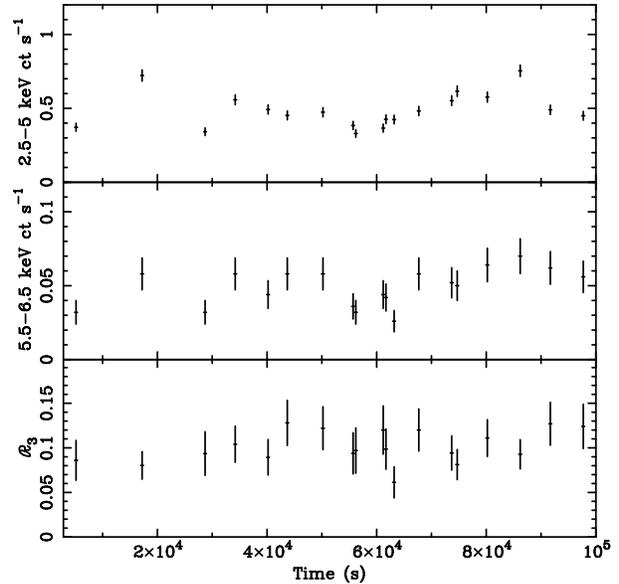

**Figure 12.** 2.5–5 keV count rate, 5.5-6.5 keV count rate and the hard to soft ratio of these two count rates ($\mathcal{R}_3$) for the August dataset. SIS0 and SIS1 data have been summed to produce these light curves. 500 s bins are used.

### 4.3  Discussion

Detailed investigation of the iron emission would require more counts in this feature than possessed by the present datasets. However, we can say the following.

i) Without the inclusion of the reflected continuum, spectral fitting with a single Gaussian gives a line which is very broad ($\sigma \sim 0.7$ keV) and strong (EW$\sim 400$ eV). Such a line would have to originate from rapidly moving material (with velocity dispersions of $\sim 30,000$ km s$^{-1}$ or more) which is overabundant in iron (George & Fabian 1991). The spectral flattening and iron K-edge introduced by the reflected continuum mimics part of the broad emission line. Fixing the reflected continuum to be from a face-on reflector subtending $2\pi$ steradians at the primary X-ray source (i.e. maximizing the reflected continuum from any disc-like geometry), the best fit line is also broad but is weaker (EW$\sim 250$ eV).

ii) Fig. 9 is suggestive of an asymmetric line, as might be expected from a relativistic accretion disk around a supermassive black hole. Additionally, the Gaussian line fits consistently produce a centroid energy less then 6.4 keV. This could result from fitting a Gaussian line to an intrinsically asymmetric emission feature. Spectral fitting using a line profile of a relativistic disk does not give a significantly better fit. Thus, longer *ASCA* observations are required in order to collect the counts necessary to define the line profile. Since this work was originally submitted, the results of such an observation have become available (Tanaka et al. 1995) and confirm the current interpretation.

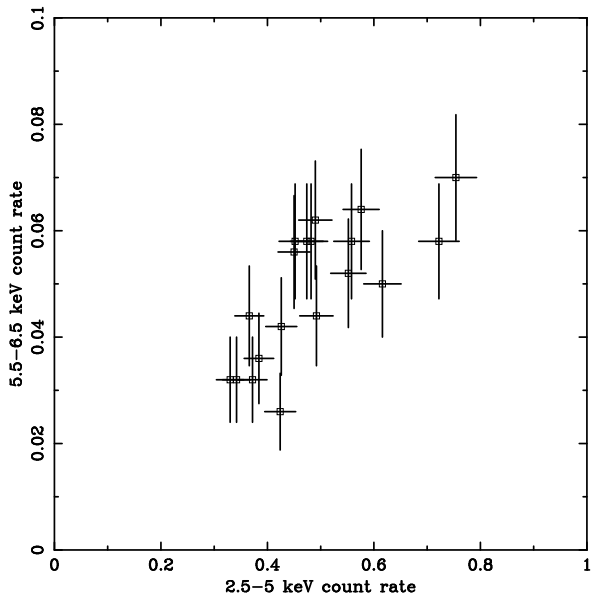

**Figure 13.** 5.5-6.5 keV count rate as a function of 2.5–5 keV count rate for the August dataset. SIS0 and SIS1 data have been summed. Each point represents 500 s of data.

iii) There are insufficient data to produce useful constraints on the variability of the iron line. Again, longer observations are required.

## 5 SUMMARY

MCG−6−30−15 is a nearby bright Seyfert 1 galaxy with strong spectral features due to X-ray reprocessing. In particular, *Ginga* shows it to have a strong reflection component (Nandra & Pounds 1994) and *ASCA* shows it to have a strong warm absorber with prominent O VII and O VIII edges in the soft X-ray spectrum. We have analysed the data from two *ASCA* PV observations of MCG−6−30−15 for spectral features and variability associated with such reprocessing.

Confirming the results of F94, we find a doubling of the warm absorbing column density from $6 \times 10^{21}$ cm$^{-2}$ to $1.1 \times 10^{22}$ cm$^{-2}$ between the three weeks separating the two *ASCA* observations. We also find a small increase in the ionization parameter $\xi$ despite a significant decrease in the ionizing flux.

We also examine short term variability of the warm absorber. We find no *flux-correlated* variability over the duration of each observation. Assuming the material is in photoionization equilibrium with the primary radiation, this implies that the material must lie between $3 \times 10^{18}$ cm and $1.5 \times 10^{19}$ cm from the source of ionizing radiation. This places it outside the BLR. However, we report the discovery of short term warm absorber variability in the July dataset. In an event lasting ∼ 10 000 s, $\tau_{\rm OVIII}$ (and, to a lesser extent, $\tau_{\rm OVII}$) undergoes a significant increase and then returns to the pre-event value. Within the context of one-zone photoionization models, the resulting upper limit on the photoionization timescale (and hence the upper limit on the distance of the warm material from the central source) is inconsistent with the lack of flux-correlated variability. We tentatively identify this event with the passage of a warm cloud across the primary X-ray source. Velocity arguments then place this cloud within the BLR. To reconcile this with the lack of observed flux-correlated variability, we would have to postulate that the material is far from equilibrium and/or that its ionization state is not controlled by photoionization by the primary X-rays. Spatial structure within the primary X-ray source together with an inhomogeneous warm absorber might also lead to complicated variability of the observed warm absorber parameters. It is interesting that in this case, where the photoionizing and corresponding absorption edges are observed simultaneously, agreement with a simple photoionizing model appears lacking.

Fluorescent cold iron emission is examined. The current data suggest that the emission line is broad (corresponding to velocities of 30 000 km s$^{-1}$ or more) and strong (EW∼ 250 eV or more). There is also evidence for asymmetry within the line. Fluorescent emission from a relativistic accretion disk is the most natural explanation of these facts. There are insufficient counts within this emission feature to usefully constrain line variability.

## ACKNOWLEDGMENTS

We thank the *ASCA* PV team for making the observations possible and Niel Brandt, Alastair Edge and Roderick Johnstone for useful discussions throughout the course of this work. CSR and PN thank PPARC, ACF thanks the Royal Society and KI thanks the JSPS and the British Council for support.

Tanaka Y. et al., 1995, Nat., in press